\documentclass[superscriptaddress,twocolumn,amsmath,amssymb,aps,prl]{revtex4-1}
\usepackage{amsmath}
\usepackage{graphicx}
\usepackage{dcolumn}
\usepackage{bm}

\begin{document}


\title{Weak localization in boron nitride encapsulated bilayer MoS$_2$} 

\author{Nikos Papadopoulos}
\email{Email: n.papadopoulos@tudelft.nl\\}
\affiliation {\small \textit Kavli Institute of Nanoscience, Delft University of Technology, Lorentzweg 1, Delft 2628 CJ, The Netherlands}
\author{Kenji Watanabe}
\affiliation {\small \textit National Institute for Materials Science, 1-1 Namiki, Tsukuba 305-0044, Japan}
\author{Takashi Taniguchi}
\affiliation {\small \textit National Institute for Materials Science, 1-1 Namiki, Tsukuba 305-0044, Japan}
\author{Herre S. J. van der Zant}
\affiliation {\small \textit Kavli Institute of Nanoscience, Delft University of Technology, Lorentzweg 1, Delft 2628 CJ, The Netherlands}
\author{Gary A. Steele}
\email{Email: g.a.steele@tudelft.nl\\}
\affiliation {\small \textit Kavli Institute of Nanoscience, Delft University of Technology, Lorentzweg 1, Delft 2628 CJ, The Netherlands}



\begin{abstract}
\noindent
We present measurements of weak localization on hexagonal boron nitride encapsulated bilayer MoS$_2$. From the analysis we obtain information regarding the phase-coherence and the spin diffusion of the electrons. We find that the encapsulation with boron nitride provides higher mobilities in the samples, and the phase-coherence shows improvement, while the spin relaxation does not exhibit any significant enhancement compared to non-encapsulated MoS$_2$. The spin relaxation time is in the order of a few picoseconds, indicating a fast intravalley spin-flip rate. Lastly, the spin-flip rate is found to be independent from electron density in the current range, which can be explained through counteracting spin-flip scattering processes based on electron-electron Coulomb scattering and extrinsic Bychkov-Rashba spin-orbit coupling. 
\end{abstract}
\maketitle

Molybdenum disulphide (MoS$_2$) is a member of the family of transition metal dichalcogenides (TMDCs) with semiconducting properties, in which the interplay between spin and other pseudo-spins, such as valley and layer index, has created new prospects for spintronics and valleytronics \cite{xu_spin_2014,schaibley_valleytronics_2016}. Bilayer MoS$_2$ is centrosymmetric and the subbands of the two K and K$^\prime$ valleys are spin degenerate under non perturbed conditions \cite{cheiwchanchamnangij_quasiparticle_2012}. When an out-of-plane electric field is applied, an inter-layer potential is generated and the inversion symmetry breaks and leading the possibility of spin-valley locking in bilayers \cite{wu_electrical_2013,lee_electrical_2016,klein_electric-field_2017}. 

Quantum corrections to the conductivity due to interference effects of charged carriers in disordered systems can provide information about fundamental properties of the carriers that reside in the system \cite{bergmann_weak_1984,kawaji_weak_1986}. They can, for example, provide information about the phase-coherence as well as about spin\cite{kawaji_weak_1986}, and other types of scattering rates of the carriers \cite{tikhonenko_weak_2008}. Specifically, in MoS$_2$ and other TMDCs, weak localization (WL) or weak antilocalization (WAL) can provide crucial information about the spin lifetime established by intravalley and intervalley scattering, as well as about the Zeeman-like splitting that is induced by the intrinsic SO coupling \cite{ochoa_spin-valley_2014,schmidt_quantum_2016}.

\begin{figure}[h]
  \begin{center}
    \includegraphics[width=7cm]{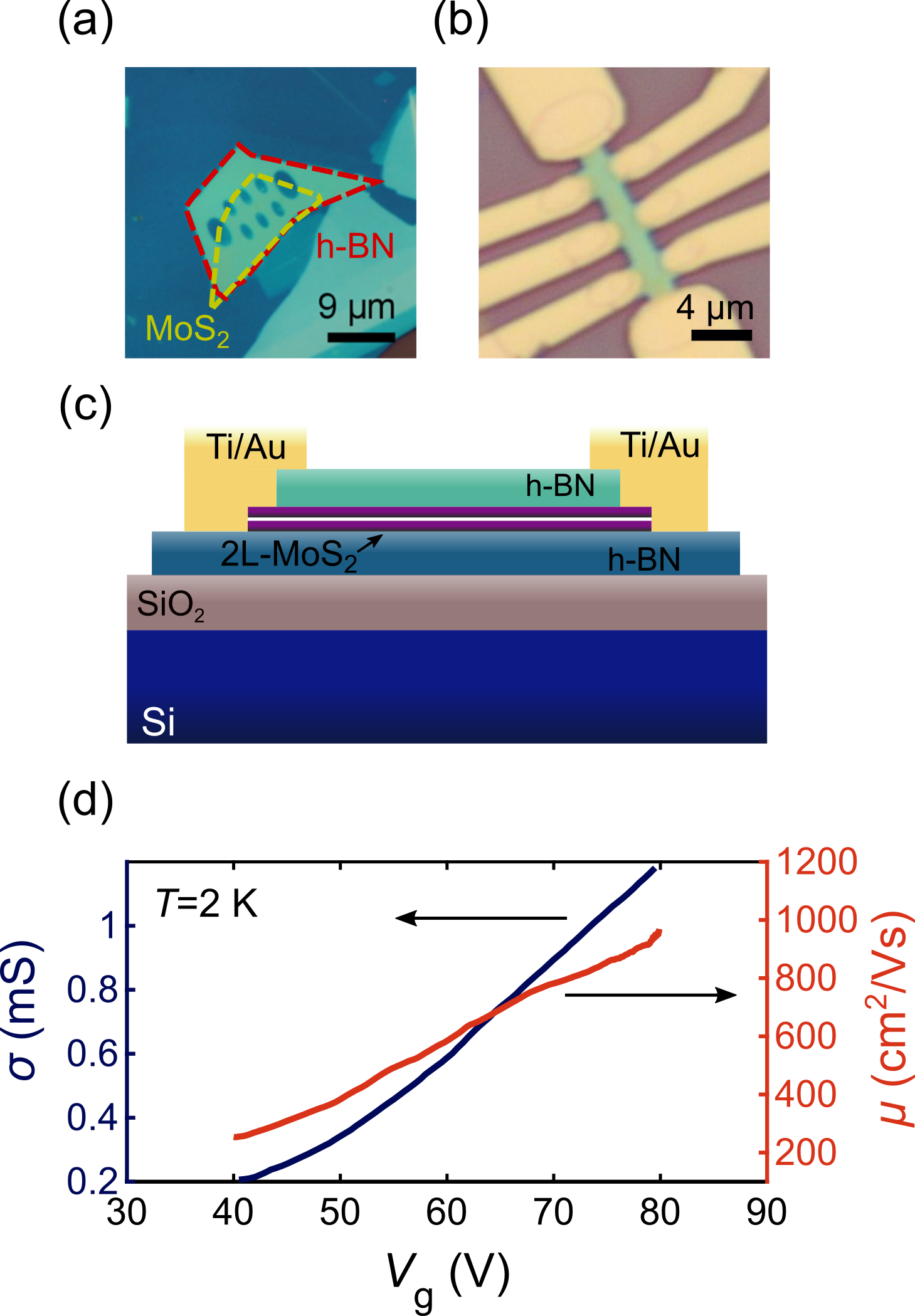}
  \end{center}
\caption{\small A high-mobility encapsulated MoS$_2$ bilayer Hall bar and device characteristics. (a) Optical image of an h-BN/2L-MoS$_2$/h-BN stack, with pre-patterned holes in the top h-BN. (b) Optical image of a completed Hall bar device. (c) Cross-sectional schematic of the device. (d) Electrical conductivity, $\sigma$, and electron mobility $\mu$ as a function of the gate voltage $V_\text{g}$, obtained from Hall measurements at $T=2$ K.}
\end{figure}

\begin{figure*}
  \begin{center}
    \includegraphics[width=17 cm]{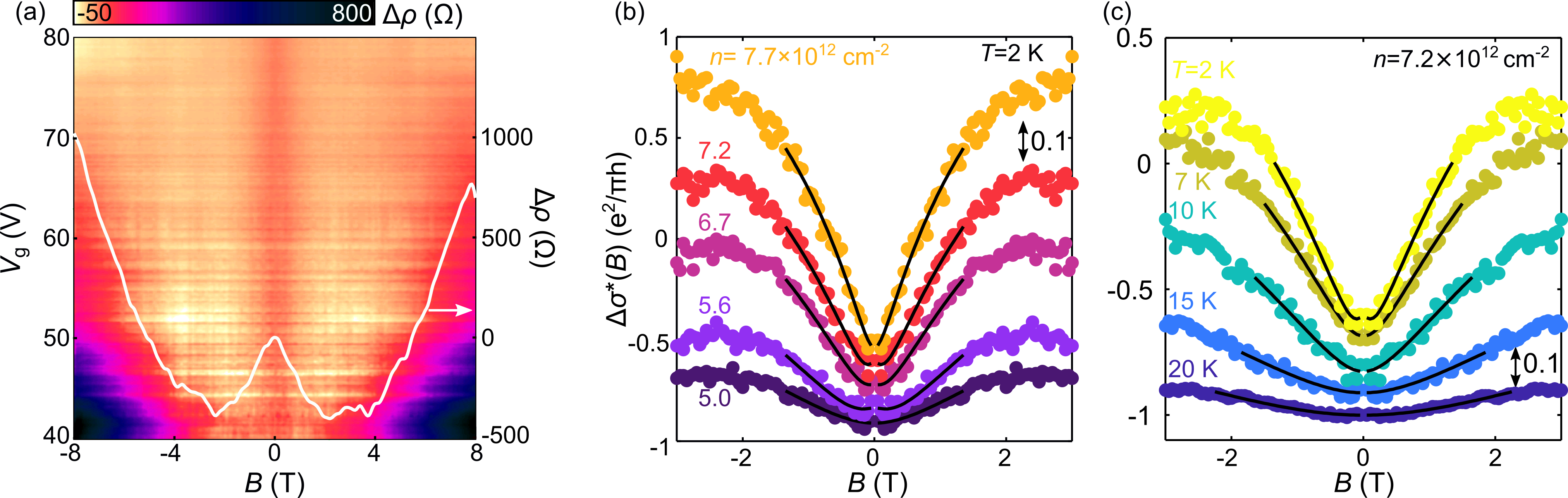}
  \end{center}
  \caption{\small Weak localization in bilayer MoS$_2$. (a) Magnetoresistivity ($\Delta\rho=\rho(B)-\rho(B=0 \text{T})$ , $\rho(B=0 \text{T})=4.56$ k${\Omega}$ at $V_\text{g}=45$ V) as a function of the back-gate voltage ($V_\text{g}$) and the magnetic field ($B$) at $T$=2 K. The overlaid linecut shows the magnetoresistivity  at a gate voltage of $V_\text{g}=45$ V. (b) Symmetrized magnetoconductivity ${\Delta\sigma^*(B)}$ as a function of magnetic field  for different electron densities $n$ measured at $T=2$ K. (c) Symmetrized magnetoconductivity as a function of magnetic field for different temperatures with $n=7.2\times10^{12}$ cm$^{-2}$. The drawn solid black lines correspond to fits using the HLN model. Data and  fitted curves have been shifted vertically by 0.1$e^2/\pi h$ for clarity.}
\end{figure*}

Although other studies have observed WL in disordered monolayer \cite{schmidt_quantum_2016} and in a few-layer MoS$_2$ samples \cite{neal_magneto-transport_2013,zhang_robustly_2017} the case of bilayer and boron nitride encapsulated has been unexplored. In this Letter, we study weak localization in high quality bilayer MoS$_2$ encapsulated in hexagonal boron nitride (h-BN). Analyzing our measurements using the Hikami-Larkin-Nagaoka (HLN) model \cite{hikami_spin-orbit_1980}, we extract the spin relaxation lengths and spin lifetimes that indicate fast spin relaxation rates through intravalley processes. Our data further suggest that the dominant source of phase-decoherence is the Altshuler-Aronov-Khmelnitsky mechanism, in which electron-electron inelastic scattering takes place \cite{altshuler_effects_1982}, similar to previous studies of quantum transport in monolayer and a few-layer MoS$_2$ \cite{neal_magneto-transport_2013,schmidt_quantum_2016}. 

Figure 1(a) and (b) show optical images of a van der Waals heterostructure (sample D2) and a final device (sample D1), respectively. To maintain the quality of MoS$_2$ during fabrication and to be able to establish good electrical contacts, we followed a different route than other studies \cite{Wang614,cui_multi-terminal_2015-1,xu_universal_2016-1}. Prior to the stacking of the heterostructure via the hot pick-up technique \cite{pizzocchero_hot_2016}, at the top h-BN sheet we opened ``windows'' in it, via standard electron beam (\textit{e}-beam) lithography followed by reactive ion etching. This allows the metallic contacts to be deposited directly on the MoS$_2$ channel \cite{wang_electronic_2015-1} and recently it was shown that this can be a good alternative to graphene contacts \cite{pisoni_gate-tunable_2018}. The ohmic behavior of the current-voltage characteristics at moderate back-gate voltages $V_\text{g}$ (Fig. S1) verifies the good quality of the contacts at low temperatures and allows the use of lock-in measurements, without the need of complicated stacking of graphene with local gates \cite{pisoni_quantized_2017,wang_electrical_2018}. In the main text we present results from data of sample D1, while in the supplemental material  data from sample D2 can be found \citep{SuppInfo}. 

The carrier density is obtained using Hall measurements and is found to be in the range of $4-8\times10^{12}$ cm$^{-2}$ for gate voltages between 40 and 80 V. The conductivity versus back-gate voltage shows typical \textit{n}-type behavior and increases when the temperature decreases due to the metallic character of the channel at these gate voltages (Fig. S1(b)). As a result of the encapsulation, the devices reach Hall mobilities of $\sim$1000 cm$^2$/Vs  at $V_\text{g}$=80 V ($n=7.9\times10^{12}$ cm$^{-2}$) (Fig. 1(d)) and field-effect mobilities of $\sim$3000 cm$^2$/Vs at $T=2$ K (Fig. S1(d)). The mean free path ($L_e$), the diffusion constant ($D$) and the momentum relaxation time ($\tau_p$) are in the range of 6-30 nm, 0.4-2.2$\times$10$^{-3}$ m$^2$s$^{-1}$ and 55-200 fs, respectively, assuming an effective mass of $0.4m_0$, where $m_0$ is the free electron mass. Also, the Fermi level ($\epsilon_F$) lies in the range of 13-24 meV above the conduction band edge. The electron mobility increases with the carrier density, which points to the presence of long-range Coulomb scattering \cite{kaasbjerg_scattering_2016}. The disorder induced doping in the MoS$_2$ channel can be obtained by extrapolating the carrier density to zero $V_\text{g}$, which gives $n_0=3.7\times10^{11}$ cm$^{-2}$.

 \begin{figure*}
  \begin{center}
    \includegraphics[width=17cm]{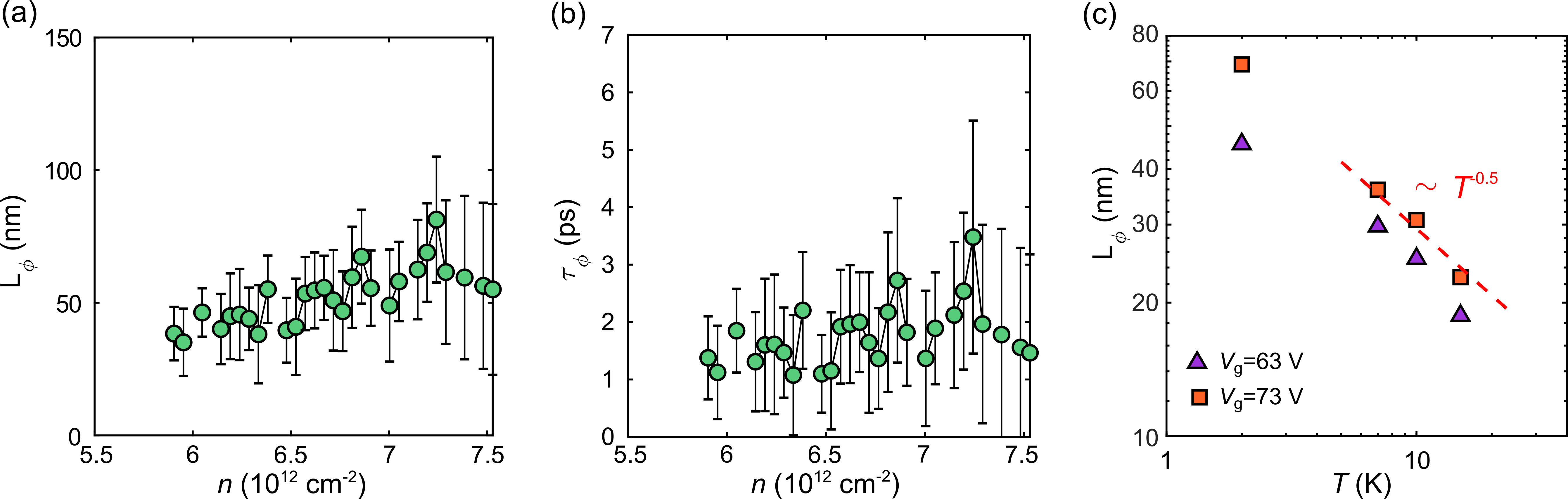}
  \end{center}
  \caption{\small Phase-coherence length (a) and phase-coherence time (b) as a function of the electron density for $T=2$ K. (c) Logarithmic plot of the phase-coherence length as a function of temperature, for two different back-gate voltages. The power law dependence with $a \sim 0.5$ suggests electron-electron scattering as the dephasing mechanism. }
\end{figure*}

At low temperatures, the magnetoresistance in our devices shows a prominent peak around $B=0$ T (Fig. 2(a)), a clear signature of weak localization of the electron wavefunctions. Figure 2(b) shows the symmetrized magnetoconductivity $\Delta\sigma^*(B)$ (where $\Delta \sigma^* (B) = (\sigma(B) + \sigma(-B))/2 - \sigma(B=0 T)$) in units of $e^2/\pi h$, for different carrier densities at $T=2$ K and in Fig. 2(c) for different temperatures at $n=7.2\times10^{12}$ cm$^{-2}$. As the carrier density increases, the dip at zero magnetic field becomes more prominent, while it declines with temperature. The former can can be attributed to an increase of the coherence length of the electrons with electron density, while the latter can be explained from a decrease in the coherence as the temperature increases. Furthermore, at high carrier densities the magnetoconductivities show oscillations that are ascribed to universal conductance fluctuations (UCFs) (see also Fig. S8). We have also observed weak localization characteristics in the sample D2, which has also similar transport characteristics (Fig. S2 \cite{SuppInfo}).

For the analysis of the low $B$-field magnetoconductivity we have employed the revised from Iordanskii \textit{et al.} theory of Hikami-Larkin-Nagaoka \cite{hikami_spin-orbit_1980, iordanskii_weak_1994, knap_weak_1996} that has been adopted for the analysis of magnetotransport in MoS$_2$ in previous reports \cite{neal_magneto-transport_2013, schmidt_quantum_2016,zhang_robustly_2017}. This model contains  spin-orbit terms, responsible for spin relaxations. We have also performed analysis with the recent theory for monolayer TMDCs \cite{ochoa_spin-valley_2014}. In the main text we focus on results based on the HLN theory. The magnetoconductivity according to the HLN model is given by \cite{hikami_spin-orbit_1980,iordanskii_weak_1994,knap_weak_1996,zhang_robustly_2017}:

\begin{multline}
{\Delta\sigma(B)}= \frac{e^2}{2\pi^{2}{\hbar}} {\times} \\ \Bigg[F(\frac{B_{\phi}+B_\text{so}}{B})+ \frac{1}{2}F(\frac{B_{\phi}+2B_\text{so}}{B})-\frac{1}{2}F(\frac{B_{\phi}}{B}) \Bigg]. 
\label{E1}
\end{multline}

\noindent
Here, $F(z)=\psi(1/2+z)-ln(z)$ and $\psi$ is the digamma function. Eq. \eqref{E1}, contains two variables: $B_{\phi}$, which corresponds to the phase-coherence field and $B_\text{so}$ which is related to the spin-orbit mediated spin relaxation processes. The black curves in Fig. 2 (b) and (c) correspond to fits with Eq. \eqref{E1}. We have limited the fitting to fields below 1.5-2.2 T so we avoid contributions from the classical magnetoresistance and from UCFs.

From the fits, we have deduced the phase-coherence length of the electrons.  In Fig. 3(a) we show the phase-coherence length as a function of the electron density for $T=2$ K, calculated from the relationship $L_{\phi}=\sqrt{ {\hbar}/(4eB_\phi) }$. The error bars have been calculated based on error propagation methods. The phase-coherence length is between 35 and 80 nm for $n=5.5-7.5 \times 10^{12}$ cm$^{-2}$, showing an increase with the density. Even though the electron density is small in comparison to other reports \cite{neal_magneto-transport_2013,schmidt_quantum_2016,zhang_robustly_2017}, the phase-coherence lengths obtained here are among the largest reported for MoS$_2$, owing to the large mobilities of the samples. Furthermore, values of $L_{\phi}$ obtained from weak localization data are in good agreement with the ones obtained from the conductance fluctuations: using the equation $\Delta B=(\hbar/e) / (\pi r)^2$ \cite{du_weak_2016} and for $\Delta B \approx 1.2-2$ T (period of oscillations), we get a length scale of 50-63 nm.  Another quantity that we obtain is the phase coherence time from the relationship: $\tau_{\phi}={L_{\phi}}^2/D$. Figure 2(b) presents the phase coherence time as a function of electron density. A weak density dependence can be observed with an  increase from $\sim$0.8 to $\sim$2 ps.  Lastly, the phase coherence length is found to depend on temperature with a power law: $L_{\phi}\propto T^{-a}$. We find values of $\alpha$ equal to 0.56 and 0.49, for  $V_\text{g}=63$ V and 73 V, respectively. Such values of $\alpha$ imply dephasing due to electron-electron scattering processes \cite{altshuler_effects_1982}, which has also been reported in graphene \cite{gorbachev_weak_2007}, black phosphorus \cite{du_weak_2016} and monolayer MoS$_2$ \cite{schmidt_quantum_2016}.

\begin{figure*}
  \begin{center}
    \includegraphics[width=17cm]{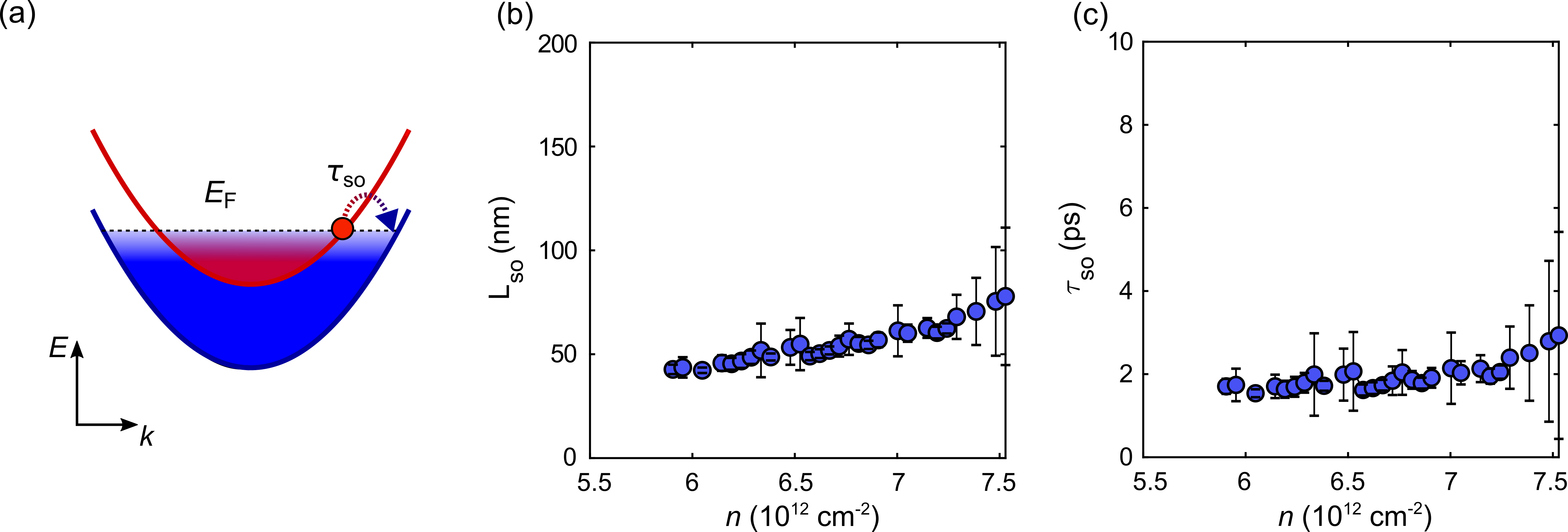}
  \end{center}
  \caption{\small Spin relaxation properties of electrons in bilayer MoS$_2$. (a) Energy-dispersion schematic illustrating spin relaxation due to intravalley spin-flip process. Different colors represent electron populations of different spin orientations. Spin relaxation length $L_\text{so}$ (b) and time $\tau_\text{so}$ (c) as a function of electron density ($T=2$ K); obtained from fitting the magnetoconductivity data to Eq. \eqref{E1}. The spin relaxation length increases with electron density, while the spin relaxation time is independent of the electron density.}
\end{figure*}

The fact that we observe weak localization in our devices indicates the absence of strong disorder that leads to intervalley spin-flip scattering and in turn to weak antilocalization \cite{ochoa_spin-valley_2014, wang_electron_2014, schmidt_quantum_2016}. Thus, the spin relaxation obtained through Eq. \eqref{E1} is mainly related to intravalley spin-flip processes (Fig. 4(a)). The dependence of the spin relaxation length ($L_\text{so}=\sqrt{ {\hbar}/(4eB_\text{so})}$) on density is presented in Fig. 4(b). The values are between 40 and 75 nm for $n=5.5-7.5 \times 10^{12}$ cm$^{-2}$, exhibiting an increase with $n$, presumably due to the increase of the diffusion constant as in the case of the phase-coherence length. The values obtained here are larger than the ones found in monolayer MoS$_2$ on SiO$_2$ (20 nm) \cite{schmidt_quantum_2016} but somewhat smaller than those obtained from a few-layer MoS$_2$ in weak localization (100-270 nm) \cite{neal_magneto-transport_2013,zhang_robustly_2017} and non local spin measurements ($\sim$ 200 nm) \cite{liang_electrical_2017}. We note that the spin relaxation and phase coherence lengths seem quite similar. Although we see no reason that these should be related, it is an interesting question if there is underlying physics behind this observation.

Unlike the spin relaxation length, the spin relaxation time is a more universal figure of merit that can be compared among different devices and materials as it does not depend on the diffusion constant. The spin relaxation time is found to be relatively fast, ${\sim}2-3$ ps (Fig. 3(b)). Recent reports on pump-probe spectroscopy on monolayer WS$_2$ have also shown fast intravalley spin-flip rates \cite{wang_intravalley_2018}. Furthermore, we find that the spin relaxation time is independent from the density. We consider two counteracting effects that can explain this observation. Firstly, according to the theoretical work of Wang \textit{et al.} \cite{wang_electron_2014} the intravalley spin-flip processes are dominated by electron-electron Coulomb scattering. As the electron density increases, the spin-flip rate should thus decrease. The spin-relaxation rate can also be tuned due to breaking of inversion symmetry in centrosymmetric TMDCs \cite{yuan_zeeman-type_2013,kormanyos_tunable_2018}. The electric-field of the back-gate can  polarize the two layers and therefore break the inversion symmetry of the system. In the case of our devices though, the inversion symmetry is already broken for the range of the back-gate voltages applied \cite{wu_electrical_2013,Lin2019} and should not affect the spin relaxation rate. The second mechanism tends to increase the spin relaxation rate through the Bychkov-Rashba SOC \cite{ochoa_spin-orbit-mediated_2013,kormanyos_spin-orbit_2014}. These two mechanisms could counteract each other resulting in a relaxation time independent on $n$. For very large electric-fields, the Bychkov-Rashba SOC dominates and the in-plane and momentum-locked effective $B$-field becomes strong enough to drive the system to WAL by spin-flip intervalley scattering \cite{yuan_zeeman-type_2013,ochoa_spin-valley_2014, yang_long-lived_2015,zhang_robustly_2017}.

In addition to the HLN model that has been typically used for the analysis the experiments of WL and WAL in MoS$_2$ \cite{neal_magneto-transport_2013,schmidt_quantum_2016,zhang_robustly_2017}, a specific model was recently developed for the analysis of of WL and WAL in monolayer TDMCs, which takes into account the interplay of the SO interaction and the multiple valleys in the band structure of TMDCs \cite{ochoa_spin-valley_2014}. The model developed by H. Ochoa \textit{et al.} and contains four free parameters ($B_{\phi}$, $B_s$, $B_e$ and $B_{\lambda}$). In the parameter range applicable to our measurements, however, we find that the parameters of the model are too strongly cross-correlated to provide a meaningful analysis of our data (see Fig. S5) We do note, however, that with similar parameters as found in the HLN fit, the model of \cite{ochoa_spin-valley_2014} does provide a theoretical prediction of the WL that is in agreement with our observations (see Supp. material for the quality of the fittings as well as results for the other fitting parameters \citep{SuppInfo}).

In summary, we have studied weak localization effects in h-BN encapsulated bilayer MoS$_2$ devices for different temperatures and electron densities. Based on the analysis of the HLN theory, we found large phase-
coherence lengths limited by to electron-electron inelastic scattering. The spin relaxation rate is found to be relatively fast and  independent from electron density. This latter observation may indicate the presence of counteracting relaxation mechanisms involving electron-electron scattering and spin-orbit interaction.

\begin{center}
  \textbf{\small ACKNOWLEDGMENTS}
\end{center}

\noindent
This work is part of the Organization for Scientific Research (NWO) and the Ministry of Education, Culture, and Science (OCW). We thank Hector Ochoa and Vladimir I. Falko for their help regarding the fitting function and the localization behavior of MoS$_2$. Growth of hexagonal boron nitride crystals was supported by the Elemental Strategy Initiative conducted by the MEXT, Japan and the CREST (JPMJCR15F3), JST.

\bibliography{WL_ref_len}

\bigskip

\end{document}